\documentclass[prl,twocolumn,showpacs,preprintnumbers,amsmath,amssymb]{revtex4}
\usepackage[dvips]{graphicx}
\usepackage{graphicx}
\usepackage{amsmath}
\usepackage{amsbsy}
\usepackage{dcolumn}
\usepackage{bm}

\newcommand{\BSCCO}{{Bi$_2$Sr$_2$CaCu$_2$O$_{8+x}$ }}

\begin{document}

\title{Structure of Bi$_2$Sr$_2$CaCu$_2$O$_{8+x}$ supermodulation from {\it ab initio} calculations}

\author{Y. He, S. Graser, P. J. Hirschfeld, and H.-P. Cheng}

\affiliation{Department of Physics, University of Florida, Gainesville, Florida 32611, USA}

\date{\today}

\begin{abstract}
We present results of density functional theory (DFT) calculation
of the structural supermodulation in Bi$_2$Sr$_2$CaCu$_2$O$_{8+x}$  structure, and
 show that the supermodulation is indeed a spontaneous symmetry
 breaking of the nominal  crystal symmetry, rather
 than a phenomenon driven by interstitial O dopants.  The structure obtained is
 in excellent quantitative agreement with recent x-ray studies, and reproduces
 several qualitative aspects of scanning tunnelling microscopy
 (STM) experiments as well.  The primary structural modulation affecting
 the CuO$_2$ plane is found to be a buckling wave of tilted
 CuO$_5$ half-octahedra, with maximum tilt angle near the phase of the
 supermodulation where recent STM experiments have
 discovered an enhancement of the superconducting gap.  We argue that the
 tilting of the half-octahedra and concommitant planar buckling
 are directly modulating the superconducting pair interaction.
\end{abstract}


\maketitle

The Bi-based family of high temperature superconducting cuprate
materials is alone  among the several families of cuprates in
possessing an incommensurate structural {\it supermodulation}. It
is not known whether this wavelike distortion of the ideal periodic crystal
structure  represents a spontaneous symmetry breaking originating
in a mismatch between the preferred bond lengths in
 the perovskite and rock salt layers of the crystal, or is driven
 by disordered interstitial oxygen atoms which dope the system.  Interest in
 this phenomenon, until recently viewed as a curiosity, has been
 reawakened by scanning tunnelling microscopy (STM)
 experiments~\cite{JASlezak:2007} which show a strong positive
 correlation of the supermodulation phase with the superconducting energy
 gap, a local measure of the strength of electron pairing.
 Theoretical calculations~\cite{BMAndersen:2007,KYYang:2007}
 suggest  that the structural modulation may couple to local
 electronic structure, which in turn influences the effective
 local pairing interaction. Knowledge of
 how the lattice distorts in the presence
 of the supermodulation could be used to calculate local
 electronic structure and electron-phonon couplings, and
 guide a theory of local pairing in the high-temperature
 superconductors.

\begin{figure}
\begin{center}
\includegraphics[width=\columnwidth]{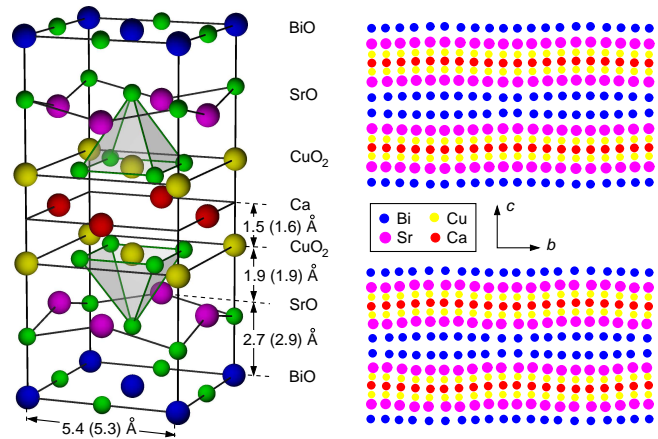}
\end{center}
\caption{Left: upper half of a nominal Bi-2212 unit cell. Lattice
constants shown are between cations in each layer, values taken
from Levin et al.~\cite{AALevin:1994} (from this work). Right: side
view of cation distribution for a layer of \BSCCO one unit cell
thick for (top) DFT bulk calculation, and (bottom) x-ray
diffraction experiment~\cite{AALevin:1994}.} \label{fig:structure}
\end{figure}

The structure of idealized \BSCCO\hskip -.1cm, with two
CuO$_2$ planes per unit cell and a critical temperature of
$T_c\approx 90K$  is shown in Fig.~\ref{fig:structure}.  The
half-cell shown is connected to an identical half-cell on the top
or bottom translated with respect to the one shown along the $a$-
or $b$-axis by half a lattice constant.
The supermodulation  is now a density wave
superposed on the structure of Fig.~\ref{fig:structure},
characterized by a wavevector ${\bf q}_{SM}$ which corresponds to
a wavelength of approximately $\lambda_{SM}=2\pi/q_{SM} \approx 26
\AA \approx 4.8~\mbox{unit cells}$.  The wavevector is
perpendicular to the wavefronts along the $a$-axis, i.e. at
45$^\circ$ with respect to the Cu-O bond direction.  Thus
superconducting \BSCCO is an almost-periodic crystal, since,
strictly speaking, no two unit cells are
identical~\cite{footnote1}.  Such structures were characterized
mathematically by de Wolff~\cite{PMdeWolff:1974}, and are known to
occur in other materials such as Na$_2$CO$_3$ as well.  The
supermodulation in \BSCCO leads to displacements of the atomic
positions of up to 0.6 \AA\, from their ideal crystalline
locations, as determined by a series of early x-ray 
diffraction~\cite{AFMarshall:1988,YLePage:1989,JMTarascon:1988,YGao:1988,VPetricek:1990}
and scanning tunnelling microscopy (STM)
experiments~\cite{MDKirk:1988}.

Historically, the \BSCCO supermodulation has been regarded as
largely irrelevant to high-temperature superconductivity because
the  material has a critical temperature of about 90K, similar to
other cuprates with two CuO$_2$ layers per unit cell which do not
display this effect.  Thus   this type of lattice distortion does
not appear to have an essential role in either enhancing or
suppressing superconductivity {\it globally}. A good deal of
attention has been paid to the effect of macroscopic distortions
of unit cell geometry  on the critical temperature when external
pressure is applied~\cite{JPLocquet:1998,GAngilella:2002,XJChen:2004}.  
In this case, however, exact changes in atomic displacements with pressure
are impossible to determine because the materials are so
complicated, even when structures are truly crystalline. On the
other hand, a recent scanning tunnelling spectroscopy (STS)
measurement~\cite{JASlezak:2007} has shown that the \BSCCO
supermodulation is correlated {\it locally} with the
 energy gap, one of the most fundamental
observables characterizing the superconducting state. By
carefully imaging the positions of Bi atoms imaged on the surface
layer of a cleaved \BSCCO crystal, the authors were able to
associate to each pixel in the topographic scan a
``supermodulation phase" $\phi_{SM}$, defining the relative
position of each Bi relative to the maximum in the incommensurate
density wave. They observed a $\sim 10$\% sinusoidal variation of
the energy gap, determined by the half of the local energy separation
of the ``coherence peaks" at positive and negative tip bias,
 with maxima generally coinciding with supermodulation
phase $\phi_{SM}=0$.

This measurement raises the intriguing possibility of a new
approach to understanding the origins of high-temperature
superconductivity: by identifying in detail the atomic
displacements which enhance superconductivity, one may be able to
determine the changes in local electronic structure or
electron-phonon coupling which modulate the pairing interaction
itself.  While  atomic-scale information is conventionally thought
to be irrelevant to superconductivity, these materials have
coherence lengths approaching the nominal unit cell size, and
there is evidence from STM studies to suggest that the pairing
interaction can indeed vary on a scale of a few \AA~\cite{TSNunner:2005a} 
in the presence of interstitial oxygen atoms.

The key missing step in this program is the precise identification
of the actual positions of the atoms present in the
supermodulation of the complex BSCCO material.  In principle, this
information can be obtained by x-ray analysis, and considerable
effort has been 
expended~\cite{XBKan:1991,AALevin:1994,DGrebille:1996,HFFan:1999,MIzquierdo:2006}
in this endeavor. The analysis is hampered by the difficulty of
analyzing superpositions of  incommensurate
harmonics~\cite{HFFan:1999}, as well as the presence of
considerable disorder in the sample, thought to consist primarily
of Bi/Sr substitutions on the
 lattice sites, as well as interstitial O
 disorder~\cite{HEisaki:2004}.  The recent STM measurements do
 provide high-resolution imaging of atomic positions, but they are
 limited to the Bi atoms on the cleaved surface.

 In this work we
 report the results of density functional theory (DFT) calculations for the
 atomic positions in the \BSCCO system including the possibility
 of a supermodulation.  While application of DFT to cuprates has been
 somewhat neglected because of its early failure to predict the
 insulating state of the parent compounds, it should produce accurate
 results for electronic states far from the Fermi level, and its success in making
 structural predictions in a variety of complex materials is well known.

{\it Method.}   Details of the application of the
 DFT method to local structure problems in \BSCCO have been given earlier in~\cite{WHC04} 
and~\cite{YHe:2006}.
The supermodulation to be studied in this work represents a wave
of small deviations from the positions represented in Fig.~\ref{fig:structure}. 
In DFT treatment of solids, a small set of
atoms is periodically repeated.  Such an approach is clearly
impossible in the case of an incommensurate supermodulation, so we
exploit the fact that the wavelength $\lambda_{SM}$ of 4.8 nominal
Bi-Bi unit cell constants of 5.4\AA\, is very close to 5 cells.
We therefore perform DFT calculations with  periodically repeated
set of 5$\times$1 or 5$\times 2$ nominal unit cells, each of
which involves 150 atoms (1420 electrons) or 300 atoms (2840
electrons), respectively.  By alternating the \BSCCO unit cell
with an insulating blocking layer in the $c$ direction, we can
also study the BiO surface for comparison to, e.g., STM
experiments. However, these calculations  were performed for
nominal {\it half}-unit cells only, due
to computer time constraints.

In addition, we study similar systems where a single dopant O
atom has been added. The lowest energy position for the dopant in the nominal
unit cell was determined in Ref.~\cite{YHe:2006}, and shown to
improve qualitatively the fit to the ARPES Fermi
surface~\cite{RMarkiewicz:2005,YHe:2006}.  In the case of a
supermodulation, the dopant may find new low energy minima in
each of the five inequivalent unit cells, and this calculation
is therefore performed independently for starting positions of the dopant
in each of them, allowing the system to relax in each case.

{\it Bulk results.}  We begin by noting that it is far from
obvious that such a calculation will find that the system, in the
absence of disorder, spontaneously deforms from its nominal
structure shown in Fig.~\ref{fig:structure} to create a
supermodulation. To our knowledge this is the first  DFT
calculation which shows effects of this type (other than the much
simpler 1D Peierls-type distortions). Since the discovery of the
supermodulation in \BSCCO, it has been controversial whether the
phenomenon is in fact an unusual spontaneous deformation of a 3D
periodic structure which lowers the total energy of the system, or
whether the interstitials necessary to dope the superconducting
material stabilize the supermodulation. Our first result is
therefore that, within DFT, the supermodulation indeed occurs
spontaneously, as shown in Fig.~\ref{fig:structure}. The waves of
displacement from the nominal crystalline positions have
significant amplitudes in all layers of the bulk system, and the
largest displacements are in the xy plane, as also found in x-ray
analyses.

\begin{figure}
\begin{center}
\includegraphics*[width=1.05\columnwidth]{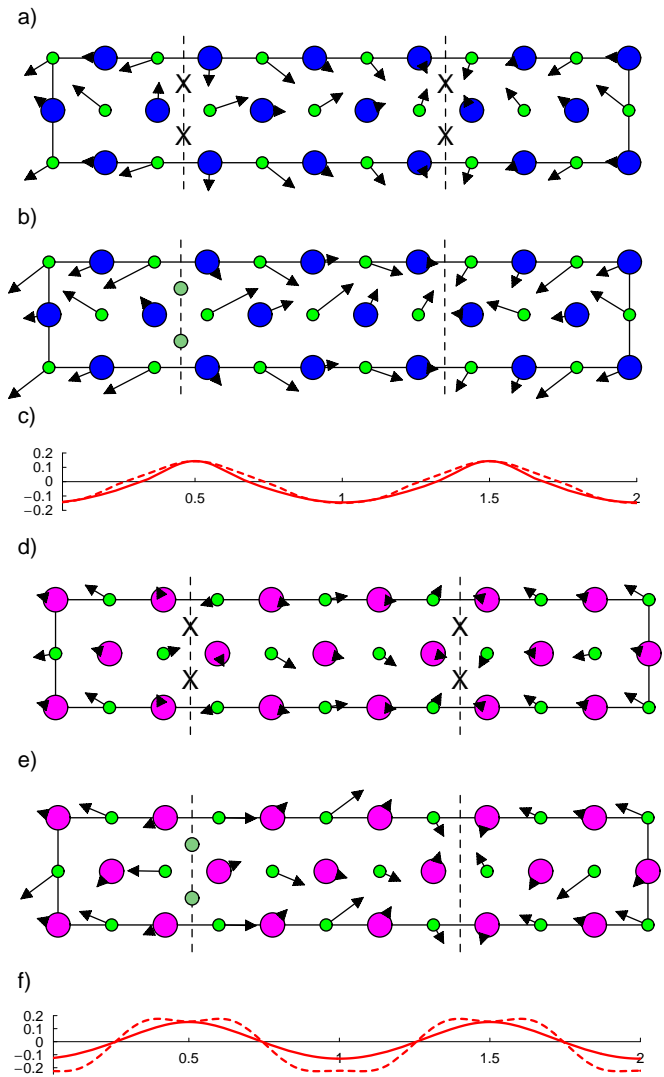}
\end{center}
\caption{Displacements in the BiO layer projected into the $x-y$
plane for a) DFT bulk calculation; b) x-ray~\cite{AALevin:1994}.
Dashed lines are mirror planes, and X's mark points of stability
found for O dopants by DFT.  Dark green circles mark positions
deduced by x-ray expt.  c) Average z-displacements of Bi for DFT
(solid) and x-ray (dashed).  d)-f): same as a)-c), but for SrO
layer. All arrow lengths were multiplied by 2 for clarity.}
\label{fig:BiO_&_SrO}
\end{figure}

In the right panel of Fig.~\ref{fig:structure}, we illustrate the
remarkable success of these calculations by comparing a side view
of the cation positions in the DFT calculation to those reported
in the x-ray measurements of Levin et al.~\cite{AALevin:1994}. The
structures are nearly indistinguishable to the eye.  Such a
representation obscures the fact that the displacements in the
$x-y$ plane from nominal crystalline positions are quite large and
more complex, however. These are displayed in Fig.~\ref{fig:BiO_&_SrO}, 
where they are indicated by arrows relative
to the nominal sites.  In the BiO layer, bulk DFT calculations and
x-ray data are again in nearly perfect agreement. On the other
hand, while certain regions of the SrO layer show quite good
agreement between experiment and theory, significant discrepancies
appear elsewhere.
This is because the atomic positions  given by the bulk DFT calculations
shown neglect the O dopant atoms, which were in fact located by Levin et al. at
the sites shown in the lower panel of Fig.~\ref{fig:BiO_&_SrO}.
The effect of these dopant interstitials on the crystal structure was
examined from the point of view of DFT both without~\cite{YHe:2006} 
and with the supermodulation present.  The local
displacements including the supermodulation are roughly as
described in Ref.~\onlinecite{YHe:2006}, but the supermodulation
is found to break the symmetry of the effective potential for the
dopant among the five nominal unit cells, leading to two effective
potential minima 0.2 eV  below other minima in the same layer,
degenerate to within 0.01eV, indicated by crosses in the figure.
Thus we expect interstitial O dopants to populate these two
different symmetry sites (shown in Fig.~\ref{fig:BiO_&_SrO})
randomly with close to equal probability.  The existence of these
two stable dopant positions is in agreement with STM~\cite{JASlezak:2007}, 
but the x-ray work~\cite{AALevin:1994}
identified only one.

 Once the existence of these dopants in the crystal is
accounted for, it is seen in Fig.~\ref{fig:BiO_&_SrO} that the
largest discrepancies in the position of the O(Sr) (apical oxygen)
between theory and experiment arise simply because the dopants
push the apical O's to the side.  The x-ray data display the
displacements of this type averaged over the random positions of
all O-dopants. The dopants do not influence the BiO layer as much
because the spontaneous supermodulation displacements move the
stoichiometric O's near the dopant sites away from those sites.

Within the framework of the DFT calculation, all qualitative
information about the all-important CuO$_2$ plane is already given
in Fig.~\ref{fig:BiO_&_SrO}.  This is because in all our
calculations the CuO$_5$ half-octahedra are always found to be
extremely rigid~\cite{YHe:2006}, and tilt as a unit when an apical
oxygen is pushed to the side as found here.  Thus significant
buckling (${\cal O}$(0.4\AA)) of the CuO$_2$ plane when the
half-octahedra tilt is also a characteristic of the
supermodulation.  Note that the x-ray calculations do not find the
same rigidity of the half-octahedra; we believe this is because
the measurement averages over both filled and unfilled dopant
sites.

\begin{figure}
\begin{center}
\includegraphics[width=.95\columnwidth]{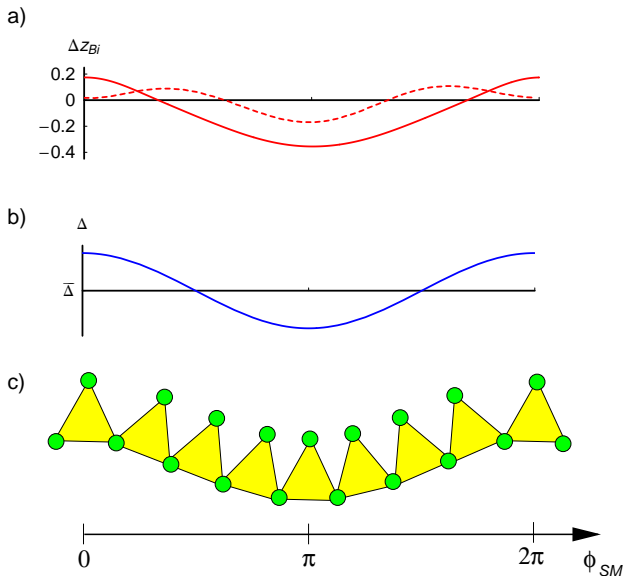}
\end{center}
\caption{a) $z$ displacements of Bi atoms in DFT surface
calculation (solid) and STM measurement~\cite{JASlezak:2007}; b)
schematic modulation of measured spectral gap vs. supermodulation
phase angle $\phi_{SM}$~\cite{JASlezak:2007}; c) schematic tilting
of CuO$_5$ half-octahedra and its dependence on $\phi_{SM}$. }
\label{fig:surface}
\end{figure}

{\it Surface results.} In the lowest energy convergent DFT result
for the free surface (``cleaved" at BiO layer) we have obtained
thus far, the supermodulation structure is qualitatively similar
to that obtained in the bulk case, with some minor surface
reconstruction. For example, the amplitude of the z-displacement
of the Bi atoms is enhanced somewhat, to about 0.57\AA ~ between
minimum and maximum, as shown in Fig.~\ref{fig:surface}.  This
pattern should now be compared to the Bi positions obtained from
STM.  The smoothed Bi z-displacements from Slezak et
al.~\cite{JASlezak:2007} are displayed for comparison in the same
Figure: while the agreement is not bad, it is clear from
examination of the topographic images in this reference and other
STM works that there are also some fundamental differences between
the $x-y$ displacements observed by the two probes, suggesting a
more significant surface reconstruction than we have been able to
reproduce by DFT.  The STM pattern is found to correspond to a
compression and rarefaction of the Bi positions, or alternatively
to a wave in the Bi-O bond angles, with no evidence for the
Bi-O-Bi chain formation discussed by Levin et al.~\cite{AALevin:1994,Alldredgeprivate}. We
have found DFT solutions of this general type, but they are higher
in energy, possibly because the half-cell calculations we have
performed do not constrain the supermodulation sufficiently. Thus
we believe a DFT description of the surface structure is possible,
but requires further work.  The goal is to complete the logical
connection between the structural supermodulation, the gap
modulation observed by STM (Fig.~\ref{fig:surface}b), and the wave
of tilted half-octahedra shown schematically in Fig.~\ref{fig:surface}c)

{\it Conclusions.} We have shown that DFT calculations with unit
cell 5 nominal unit cells long describe
quantitatively the displacements of atoms in the bulk of  the
\BSCCO material with its well-known supermodulation. To our
knowledge this is the first ab initio calculation able to
accurately predict a 3D spontaneous structural deformation of this
type, and it provides strong evidence that the supermodulation is
a spontaneous breaking of crystal symmetry rather than being
driven by interstitial defects. Calculations for a free BiO
surface agree qualitatively in some respects with STM
measurements, but differ in significant ways which we do not
understand at this time.  All calculations indicate, however, that
the crucial structural element modulated in the 2212 case is the
CuO$_5$ half-octahedra; these objects deform rigidly and in such a
way that the maximum tilt angle corresponds to the maximum gap
observed by STM.  This conclusion is identical to one we reached
earlier in the case of O dopant atoms.  The fact that two
independent perturbations of the crystal lead to a) similar
distortions of the half-octahedra, and b) similar local
enhancements of the superconducting gap, strongly suggests that
the electronic structure changes caused by these tilts are
directly related to the local pair interaction.  The obvious next
step is to downfold this information to obtain local variations in
hopping matrix elements, superexchange constants, and
electron-phonon couplings.  This work is in progress.

{\it Acknowledgements.} The authors are grateful to
J. Alldredge, B.M. Andersen, J. C. Davis, T. Devereaux, and J.A. Slezak
for stimulating discussions.
Calculations are done on DOE/NERSC Center and the High Performance 
Computer Center at UF. Allocations at NERSC are from a DOE/BES user 
grant and a user project at CNMS/ORNL. Partial funding was
provided by DOE Grants No. DE-FG02-05ER46236 (PJH), No.
DE-FG02-97ER45660 (H-PC), and No. DE-FG02-02ER45995 (H-PC), as
well as the NSF/DMR/ ITR-medium program under Contract No.
DMR-032553.


\begin{thebibliography}{99}
\bibitem{footnote1} We follow convention in referring to the nominal ``unit
cell" of the BSCCO crystal as being the approximate repeating unit
containing 8 Bi atoms.
\end{thebibliography}
\end{document}